\documentclass[12pt,a4paper]{article}

\usepackage{latexsym,amssymb,amsmath,amsbsy,epsfig}

\newcommand{\be}{\begin{equation}}
\newcommand{\ee}{\end{equation}}
\newcommand{\ba}{\begin{eqnarray}}
\newcommand{\ea}{\end{eqnarray}}
\newcommand{\bs}{\begin{subequations}}
\newcommand{\es}{\end{subequations}}
\newcommand{\no}{\nonumber\\}

\newcommand{\pmns}{U}
\newcommand{\mnu}{\mathcal{M}_\nu}

\newcommand{\diag}{\mbox{diag}}

\begin{document}
\renewcommand{\thefootnote}{\fnsymbol{footnote}}

\title{
\normalsize \hfill CFTP/14-018
\\*[7mm]
\LARGE New textures for the lepton mass matrices}

\author{
P.\ M.\ Ferreira$^{(1,2)}$\thanks{\tt pmmferreira@fc.ul.pt} \
and L.\ Lavoura$^{(3)}$\thanks{\tt balio@cftp.tecnico.ulisboa.pt} \
\\[5mm]
$^{(1)} \! $
\small Instituto Superior de Engenharia de Lisboa \\
\small 1959-007 Lisboa, Portugal
\\[1mm]
$^{(2)} \! $
\small Centro de F\'\i sica Te\'orica e Computacional, Universidade de Lisboa \\
\small 1649-003 Lisboa, Portugal
\\[1mm]
$^{(3)} \! $
\small Universidade de Lisboa, Instituto Superior T\'ecnico, CFTP \\
\small 1049-001 Lisboa, Portugal
\\[5mm]
}

\date{2 December 2014}

\maketitle

\begin{abstract}
We study predictive textures for the lepton mass matrices
in which the charged-lepton mass matrix
has either four or five zero matrix elements
while the neutrino Majorana mass matrix has,
respectively,
either four or three zero matrix elements.
We find that all the viable textures of these two kinds
share many predictions:
the neutrino mass spectrum is inverted,
the sum of the light-neutrino masses is close to 0.1~eV,
the Dirac phase $\delta$ in the lepton mixing matrix
is close to either $0$ or $\pi$,
and the mass term responsible for neutrinoless double-beta decay
lies in between 12 and 22~meV.
\end{abstract}

\newpage

\renewcommand{\thefootnote}{\arabic{footnote}}

\section{Introduction}

The origin of neutrino masses,
the reasons behind their smallness,
and the structure of lepton mixing
are still unanswered questions.
There has been a great deal of theoretical work in this area,
trying to provide answers based on such diverse ideas as,
for instance,
seesaw mechanisms,
radiative generation of the neutrino masses,
Abelian and non-Abelian symmetries imposed on the leptonic sector,
and `textures' for the leptonic mass matrices.
In the past few years,
a wealth of experimental data concerning neutrino oscillations---in particular
the recent confirmation~\cite{Abe:2011fz,An:2012eh,Ahn:2012nd}
of a non-zero value for the mixing angle $\theta_{13}$---became available,
allowing theorists to test their models
and discard those that do not conform to the experimental discoveries.
Here,
we shall consider new textures for the leptonic mass matrices
and investigate what the most recent and stringent phenomenological data
say about them and their predictive power.

In this paper we work in the context of a model with three light neutrinos
which are Majorana particles.
The lepton mass terms are given by
\be
\mathcal{L}_\mathrm{mass} =
- \bar \ell_L M_\ell \ell_R
- \bar \ell_R M_\ell^\dagger \ell_L
+ \frac{1}{2} \left( \nu^T C^{-1} M_\nu \nu
- \bar \nu M_\nu^\ast C \bar \nu^T \right),
\ee
where $C$ is the charge-conjugation matrix in Dirac space.
The three light-neutrino fields in the column-vector $\nu$ are left-handed.
The neutrino mass matrix $M_\nu$ acts in flavour space and is symmetric.
Let the two mass matrices be bi-diagonalized as
\bs
\ba
\label{ULL}
U_L^\dagger M_\ell U_R &=& \diag \left( m_e,\ m_\mu,\ m_\tau \right),
\\
\label{Unu}
U_\nu^T M_\nu U_\nu &=& \diag \left( m_1,\ m_2,\ m_3 \right),
\ea
\es
where $U_L$,
$U_R$,
and $U_\nu$ are $3 \times 3$ unitary matrices.
Then,
the lepton mixing matrix is
\be
\label{pmns}
\pmns = U_L^\dagger U_\nu.
\ee
Even though $M_\ell$ is more fundamental,
in practice we only need to consider
\be
H \equiv M_\ell M_\ell^\dagger,
\ee
since it is its diagonalization that fixes the matrix $U_L$
which appears in $\pmns$:
\be
\label{bkpdl}
U_L^\dagger H U_L = \diag \left( m_e^2,\ m_\mu^2,\ m_\tau^2 \right).
\ee
Let $\mathcal{M}_\nu$ denote the neutrino mass matrix
in the basis where the charged-lepton mass matrix is diagonal.
Then,
\be
\mathcal{M}_\nu = U_L^T M_\nu U_L.
\ee

There have been many attempts at using non-Abelian symmetries
to constrain lepton mixing~\cite{Lam:2008rs}--\cite{Lavoura:2014kwa}.
This is usually done with the goal of obtaining `mass-independent schemes',
wherein the constraints on $\pmns$
do not depend on the values of the lepton masses.
However,
those attempts appear to have reached their limits~\cite{Fonseca:2014koa}.
A simpler avenue,
at least in group-theoretical terms,
is provided by Abelian symmetries.
In appropriate bases for the lepton and Higgs fields,
they enforce `texture zeros' in the lepton mass matrices,
but they cannot enforce relationships among their nonzero matrix elements.
In the pioneering work of ref.~\cite{Frampton:2002yf},
$M_\ell$ was assumed to be diagonal,
hence to have six zero matrix elements,
while $M_\nu$ had two zero matrix elements.
This was later generalized to the situation wherein
$M_\ell$ is diagonal and $M_\nu^{-1}$ has
two zero matrix elements~\cite{Lavoura:2004tu};
mixed situations in which both $M_\nu$ and $M_\nu^{-1}$
have one zero matrix element,
while $M_\ell$ remains diagonal,
were considered in ref.~\cite{Dev:2010if}.

In this work we propose new textures for the lepton mass matrices
which are in principle as predictive
as the ones considered in refs.~\cite{Frampton:2002yf,
Lavoura:2004tu,
Dev:2010if}.
Let $(m, n)$ denote a class of textures with
$m$ nonzero matrix elements in $M_\ell$
and $n$ nonzero matrix elements in $M_\nu$.\footnote{$M_\nu$ is symmetric
because it is a Majorana mass matrix.
Hence,
only six out of its nine matrix elements are independent.
The integer $n$ denotes the number of \emph{independent}\/ matrix elements
which do not vanish;
if some of those elements are off-diagonal,
then the actual number of nonzero entries in $M_\nu$ is larger than $n$.}
Then,
the textures mentioned at the end of the previous paragraph
are in the $(3, 4)$ class.
In this paper we consider predictive,
viable textures in the $(4, 3)$ and $(5, 2)$ classes.
Those textures are in principle just as predictive
as the ones in class $(3, 4)$;
each of them has  \emph{eight}\/ degrees of freedom---seven moduli
and one rephasing-invariant phase---in the matrices $H$ and $M_\nu$.
Those eight degrees of freedom are meant
to fit \emph{ten}\/ observables---the three charged-lepton masses
$m_{e, \mu, \tau}$,
the three neutrino masses $m_{1, 2, 3}$,
the three lepton mixing angles $\theta_{12, 13, 23}$,
and the Dirac phase $\delta$.
(We do not care about the Majorana phases in $\pmns$
because they are not observable in neutrino oscillations.
However,
we shall specify the predictions of our textures for the mass term
responsible for neutrinoless double-beta decay,
$m_{\beta \beta} \equiv \left| \left( \mnu \right)_{ee} \right|$.)
So,
in principle each texture yields \emph{two}\/ predictions,
which may conveniently be taken to be
one prediction for the overall scale of the neutrino masses
and one prediction for $\cos{\delta}$.

It has long been known~\cite{Grimus:2004hf}
that \emph{any}\/ mass-matrix texture,
in particular any set of matrices $M_\ell$ and $M_\nu$
with some zero matrix elements,
can be implemented in a suitable extension
of the Standard Model of the electroweak interactions,
furnished with both additional scalar multiplets
and appropriate Abelian symmetries.
We rely on this fact to assert that all the textures in this paper
may be implemented in renormalizable models.
However,
we shall not attempt here
to construct a specific model for any of the textures;
we also do not attempt to search
for the simplest model which might justify
any given texture~\cite{Felipe:2014vka}.

We emphasize that all the textures will be analyzed in this paper
only at the `classical' level,
\textit{i.e.}\ we shall neglect both quantum corrections to the mass matrices
and renormalization-group effects.

The texture-zero approach for the mass matrices
pursued in this paper is inherently limited in its scope and objectives.
Even if it were found that
the experimental data fully agree with the predictions of some texture,
we would not be sure that
the mass matrices indeed have that texture,
because there are many sets of mass matrices
leading to the same observables---in particular,
any two sets of mass matrices connected among themselves
through a weak-basis transformation lead to the same observables.
Further studies would be necessary in order
to identify specific models that lead to mass matrices with that texture
and also to identify other observable predictions of those models,
\textit{viz.}\ extra particles and interactions that they may feature.
So,
the study of textures may be looked upon as just the first part
of a longer search for models of `new physics'.
Still,
that study has some relevance in itself,
since it may suggest the most likely ranges for some observables---for instance,
knowing whether the phase $\delta$ is more likely to be large
or small---and which correlations among observables may be expected
and are enforceable through well-defined renormalizable models.

This paper is organized as follows.
In section~2 we derive all the viable (5, 2) textures
and briefly survey their predictions.
We do the same for (4, 3) textures in section~3.
A listing of all the viable textures that we have found,
and a summary of their predictions,
is provided in section~4.

\section{(5, 2) textures}

Since all three charged leptons are massive,
the determinant of $M_\ell$ cannot vanish.
Therefore,
through an appropriate permutation of the columns of $M_\ell$---this
permutation changes $U_R$ but does not change $U_L$,
hence it leaves $\pmns$ invariant---one may always obtain
the $\left( 1,\ 1 \right)$,
$\left( 2,\ 2 \right)$,
and $\left( 3,\ 3 \right)$ matrix elements of $M_\ell$ to be nonzero.
Since in a (5, 2) texture $M_\ell$ has five nonzero matrix elements,
there are then $\left( 6 \times 5 \right) /\, 2 = 15$ possibilities:
\bs
\label{ivuro}
\ba
M_\ell &\sim& \left( \begin{array}{ccc}
\times & \times & 0 \\ \times & \times & 0 \\ 0 & 0 & \times
\end{array} \right),
\\
M_\ell &\sim& \left( \begin{array}{ccc}
\times & 0 & \times \\ 0 & \times & 0 \\ \times & 0 & \times
\end{array} \right),
\\
M_\ell &\sim& \left( \begin{array}{ccc}
\times & 0 & 0 \\ 0 & \times & \times \\ 0 & \times & \times
\end{array} \right);
\ea
\es
\bs
\label{viogp}
\ba
\label{re4}
M_\ell &\sim& \left( \begin{array}{ccc}
\times & 0 & \times \\ 0 & \times & 0 \\ 0 & \times & \times
\end{array} \right),
\\
M_\ell &\sim& \left( \begin{array}{ccc}
\times & 0 & 0 \\ 0 & \times & \times \\ \times & 0 & \times
\end{array} \right),
\\
M_\ell &\sim& \left( \begin{array}{ccc}
\times & 0 & 0 \\ 0 & \times & 0 \\ \times & \times & \times
\end{array} \right);
\ea
\es
\bs
\label{vjipd}
\ba
M_\ell &\sim& \left( \begin{array}{ccc}
\times & \times & 0 \\ 0 & \times & \times \\ 0 & 0 & \times
\end{array} \right),
\\
M_\ell &\sim& \left( \begin{array}{ccc}
\times & 0 & 0 \\ \times & \times & 0 \\ 0 & \times & \times
\end{array} \right),
\\
M_\ell &\sim& \left( \begin{array}{ccc}
\times & 0 & 0 \\ \times & \times & \times \\ 0 & 0 & \times
\end{array} \right);
\ea
\es
\bs
\label{refos}
\ba
M_\ell &\sim& \left( \begin{array}{ccc}
\times & \times & 0 \\ 0 & \times & 0 \\ \times & 0 & \times
\end{array} \right),
\\
M_\ell &\sim& \left( \begin{array}{ccc}
\times & 0 & \times \\ \times & \times & 0 \\ 0 & 0 & \times
\end{array} \right),
\\
M_\ell &\sim& \left( \begin{array}{ccc}
\times & \times & \times \\ 0 & \times & 0 \\ 0 & 0 & \times
\end{array} \right);
\ea
\es
\ba
\label{re1}
M_\ell &\sim& \left( \begin{array}{ccc}
\times & 0 & 0 \\ \times & \times & 0 \\ \times & 0 & \times
\end{array} \right);
\\
\label{re2}
M_\ell &\sim& \left( \begin{array}{ccc}
\times & \times & 0 \\ 0 & \times & 0 \\ 0 & \times & \times
\end{array} \right);
\\
\label{re3}
M_\ell &\sim& \left( \begin{array}{ccc}
\times & 0 & \times \\ 0 & \times & \times \\ 0 & 0 & \times
\end{array} \right).
\ea
Equations~(\ref{viogp}) are equivalent\footnote{It may easily
be demonstated that,
through unitary redefinitions of the right-handed charged leptons,
one may transform any one of eqs.~(\ref{viogp}) into any other of them.}
because they all lead to $H_{12} = 0$,
hence to the same constraints on $U_L$ and on $\pmns$.
Similarly,
eqs.~(\ref{vjipd}) feature $H_{13} = 0$
and eqs.~(\ref{refos}) have $H_{23} = 0$.
Also,
$\left( H^{-1} \right)_{23} = 0$ for eq.~(\ref{re1}),
$\left( H^{-1} \right)_{13} = 0$ for eq.~(\ref{re2}),
and $\left( H^{-1} \right)_{12} = 0$ for eq.~(\ref{re3}).

In a (5, 2) texture $M_\nu$ has only two nonzero matrix elements.
Leaving aside possible reorderings of the rows and columns of $M_\nu$,
there are only four possibilities:
\bs
\ba
\label{1x}
M_\nu &\sim& \left( \begin{array}{ccc}
0&\times&\times \\ \times&0&0 \\ \times&0&0
\end{array} \right),
\\
\label{2}
M_\nu &\sim& \left( \begin{array}{ccc}
\times&0&0 \\ 0&\times&0 \\ 0&0&0
\end{array} \right),
\\
\label{2x}
M_\nu &\sim& \left( \begin{array}{ccc}
0&\times&0 \\ \times&0&0 \\ 0&0&\times
\end{array} \right),
\\
\label{3}
M_\nu &\sim& \left( \begin{array}{ccc}
0&0&0 \\ 0&0&\times \\ 0&\times&\times
\end{array} \right).
\ea
\es
Both eqs.~(\ref{1x}) and~(\ref{2x}) lead to two degenerate neutrinos
and are therefore incompatible with experiment.

With eq.~(\ref{2}) lepton mixing originates fully in $M_\ell$;
indeed,
one then has
$\pmns = U_L^\dagger$ but for possible reorderings of the columns of $\pmns$.
For two physical neutrinos $\nu_i$ and $\nu_j$ with $i \neq j$,
\ba
\label{xi1}
H_{ij} &=&
m_e^2 U_{ei}^\ast U_{ej}
+ m_\mu^2 U_{\mu i}^\ast U_{\mu j}
+ m_\tau^2 U_{\tau i}^\ast U_{\tau j}
\no &=&
\left( m_\mu^2 - m_e^2 \right) U_{\mu i}^\ast U_{\mu j}
+ \left( m_\tau^2 - m_e^2 \right) U_{\tau i}^\ast U_{\tau j}.
\ea
So,
\be
\label{xi2}
H_{ij} = 0\, \Rightarrow\,
- \frac{U_{\tau i}^\ast U_{\tau j}}{U_{\mu i}^\ast U_{\mu j}} =
\frac{m_\mu^2 - m_e^2}{m_\tau^2 - m_e^2}
\approx
\frac{m_\mu^2}{m_\tau^2}
\approx \frac{1}{280}.
\ee
Similarly,
\be
\label{xi3}
\left( H^{-1} \right)_{ij} = 0\, \Rightarrow\,
- \frac{U_{ei}^\ast U_{ej}}{U_{\mu i}^\ast U_{\mu j}} =
\frac{m_\mu^{-2} - m_\tau^{-2}}{m_e^{-2} - m_\tau^{-2}}
\approx
\frac{m_e^2}{m_\mu^2}
\approx \frac{1}{43000}.
\ee
Phenomenologically,
there are no two columns $i$ and $j$ of $U$ such that either
$\left| \left( U_{\tau i} U_{\tau j} \right)
/ \left( U_{\mu i} U_{\mu j} \right) \right|$
or
$\left| \left( U_{ei} U_{ej} \right)
/ \left( U_{\mu i} U_{\mu j} \right) \right|$
are allowed to be so much smaller than unity
as indicated by eqs.~(\ref{xi2}) and~(\ref{xi3}).
Therefore,
with eq.~(\ref{2}) either $H_{ij} = 0$
or $\left( H^{-1} \right)_{ij} = 0$ are phenomenologically forbidden
for $i \neq j$.
If,
together with eq.~(\ref{2}),
the form of $M_\ell$ is as in one of eqs.~(\ref{ivuro}),
then lepton mixing would only be $2 \times 2$,
which is also incompatible with experiment.
Therefore,
eq.~(\ref{2}) must be excluded,
just as eqs.~(\ref{1x}) and~(\ref{2x}).

We shall therefore concentrate on eq.~(\ref{3}).
With that form for $M_\nu$,
\emph{one neutrino is massless;
this is one of the predictions of viable (5, 2) textures}.

If $M_\nu$ is as in eq.~(\ref{3})
while $M_\ell$ is as in one of eqs.~(\ref{ivuro}),
then the matrix $\pmns$ has one vanishing matrix element.
This contradicts the phenomenology.
Therefore,
we may exclude eqs.~(\ref{ivuro})
and concentrate exclusively on the other possibilities for $M_\ell$.
As we have seen,
they can be subsumed in six different possibilities:
$H_{12} = 0$,
$H_{13} = 0$,
$H_{23} = 0$,
$\left( H^{-1} \right)_{23} = 0$,
$\left( H^{-1} \right)_{13} = 0$,
and $\left( H^{-1} \right)_{12} = 0$.

Let
\be
A \equiv \left( \begin{array}{ccc}
0&1&0\\1&0&0\\0&0&1
\end{array} \right),
\quad
B \equiv \left( \begin{array}{ccc}
0&0&1\\0&1&0\\1&0&0
\end{array} \right).
\ee
Then,
the permutation group of three objects is represented by
\be
S_3 = \left\{
A,\ B,\ A B A,\ A B,\ B A,\ A^2
\right\}.
\label{ss33}
\ee
Let $Z$ be any of the six matrices in $S_3$.
Those matrices are orthogonal,
hence $Z^{-1} = Z^T$.
Interchanging the rows and columns of $M_\nu$
is equivalent to making $M_\nu \to Z M_\nu Z^T$.
But $U_\nu \to Z U_\nu$ when this happens.
Therefore $\pmns \to U_L^\dagger Z U_\nu$.
This is equivalent to letting $U_L \to Z^\dagger U_L$ or $H \to Z^\dagger H Z$,
which means a reordering of the rows and columns of $H$.

So,
a reordering of the rows and columns of $M_\nu$
is equivalent to an analogous reordering of the rows and columns of $H$.
Therefore,
instead of considering separately each of the three conditions $H_{12} = 0$,
$H_{13} = 0$,
and $H_{23} = 0$,
one may consider only the condition $H_{12} = 0$ provided one allows
for all the possible reorderings of the rows and columns of $M_\nu$.
We thus conclude that there are twelve potentially viable (5, 2) textures:

\bs
\label{mxopd}
\ba
H_{12} = 0 & \mathrm{and} &
M_\nu \sim \left( \begin{array}{ccc}
0&0&0 \\ 0&0&\times \\ 0&\times&\times
\end{array} \right),
\\
H_{12} = 0 & \mathrm{and} &
M_\nu \sim \left( \begin{array}{ccc}
0&0&\times \\ 0&0&0 \\ \times&0&\times
\end{array} \right),
\\
H_{12} = 0 & \mathrm{and} &
M_\nu \sim \left( \begin{array}{ccc}
0&0&0 \\ 0&\times&\times \\ 0&\times&0
\end{array} \right),
\\
H_{12} = 0 & \mathrm{and} &
M_\nu \sim \left( \begin{array}{ccc}
0&\times&0 \\ \times&\times&0 \\ 0&0&0
\end{array} \right),
\\
H_{12} = 0 & \mathrm{and} &
M_\nu \sim \left( \begin{array}{ccc}
\times&\times&0 \\ \times&0&0 \\ 0&0&0
\end{array} \right),
\\
H_{12} = 0 & \mathrm{and} &
M_\nu \sim \left( \begin{array}{ccc}
\times&0&\times \\ 0&0&0 \\ \times&0&0
\end{array} \right),
\ea
\es
together with the six textures that result from substituting $H_{12} = 0$
by $\left( H^{-1} \right)_{12} = 0$ in each of eqs.~(\ref{mxopd}).

With either $H_{12} = 0$ or $\left( H^{-1} \right)_{12} = 0$,
the matrix $H$ can always be made real through a rephasing,
\textit{i.e.}\ there is always a diagonal unitary matrix $Y_\ell$ such that
\be
Y_\ell^\ast H Y_\ell = H_\mathrm{real}
\ee
has real matrix elements.
The real and symmetric matrix $H_\mathrm{real}$
is diagonalized by an orthogonal matrix $O_\ell$:
\be
\label{nbkpd}
O_\ell^T H_\mathrm{real} O_\ell = \diag \left( m_e^2,\ m_\mu^2,\ m_\tau^2 \right).
\ee
Since either $H_\mathrm{real}$ or its inverse has one vanishing matrix element,
it contains only five degrees of freedom;
three of them correspond to the three charged-lepton masses
and the remaining two are implicitly contained in $O_\ell$.
Thus,
$O_\ell$ is not fully general---a general
$3 \times 3$ orthogonal matrix has three degrees of freedom,
not just two.

Similarly,
phases may be withdrawn from the matrix $M_\nu$ in eq.~(\ref{3}):
\be
\label{bjpsr}
Y_\nu \left( \begin{array}{ccc} 0 & 0 & 0 \\ 0 & 0 & f e^{i \phi} \\
0 & f e^{i \phi} & r e^{i \rho}  \end{array} \right) Y_\nu =
\left( \begin{array}{ccc}
0 & 0 & 0 \\ 0 & 0 & f \\ 0 & f & r \end{array} \right),
\ee
where $f$ and $r$ are non-negative real and $Y_\nu = \diag \left( e^{i \xi},\
e^{i \left( \rho / 2 - \phi \right)},\  e^{- i \rho / 2} \right)$,
the phase $\xi$ being arbitrary.
Therefore,
the lepton mixing matrix always ends up being
\be
\label{ob}
\pmns = O_\ell^T X O_\mathrm{b},
\ee
where $O_\mathrm{b}$ is the real,
orthogonal matrix that diagonalizes the real version of $M_\nu$
while $X$ is a diagonal unitary matrix containing \emph{only one}\/ phase.
This is because the arbitrariness of the phase $\xi$ in $Y_\nu$
allows one to absorb one phase in $X$.

Let us define
\be
\delta \equiv \Delta m^2_\mathrm{sol} \equiv m_2^2 - m_1^2,
\quad
\Delta \equiv \Delta m^2_\mathrm{atm} \equiv \left| m_3^2 - m_1^2 \right|,
\quad
\varepsilon \equiv \frac{\delta}{\Delta} \approx \frac{1}{30}.
\ee
With a massless neutrino
there are two possibilities for the neutrino mass spectrum:
either it is `normal'
(which we call ``case n''),
and then\footnote{We use in this paper
the quantity $\sqrt{\Delta} \approx 0.5$~eV
as the unit for all neutrino masses.}
\be
\label{jbipd}
\frac{m_1}{\sqrt{\Delta}} = 0, \quad
\frac{m_2}{\sqrt{\Delta}} = \sqrt{\varepsilon}, \quad
\frac{m_3}{\sqrt{\Delta}} = 1, \quad
\frac{m_1 + m_2 + m_3}{\sqrt{\Delta}} = 1 + \sqrt{\varepsilon};
\ee
or it is `inverted'
(which we call ``case i''),
and then
\be
\frac{m_1}{\sqrt{\Delta}} = 1, \quad
\frac{m_2}{\sqrt{\Delta}} = \sqrt{1 + \varepsilon}, \quad
\frac{m_3}{\sqrt{\Delta}} = 0, \quad
\frac{m_1 + m_2 + m_3}{\sqrt{\Delta}} = 1 + \sqrt{1 + \varepsilon}.
\ee
Notice that
\bs
\label{vkpst}
\ba
\frac{m_1 + m_2 + m_3}{\sqrt{\Delta}} &\approx& 1 \quad
\mathrm{in\ case\ n,\ but}
\\
\frac{m_1 + m_2 + m_3}{\sqrt{\Delta}} &\approx& 2 \quad
\mathrm{in\ case\ i}.
\ea
\es

Suppose the initial $M_\nu$ was as in eq.~(\ref{3}).
Then,
after withdrawing phases from it,
we would have,
in case n
\be
M_\nu \to M_\mathrm{n} = \sqrt{\Delta} \left( \begin{array}{ccc} 0 & 0 & 0 \\
0 & 0 & \varepsilon^{1/4} \\*[1mm] 0 & \varepsilon^{1/4} & 1 - \sqrt{\varepsilon}
\end{array} \right),
\ee
while,
in case i
\be
M_\nu \to M_\mathrm{i} = \sqrt{\Delta} \left( \begin{array}{ccc} 0 & 0 & 0 \\
0 & 0 & \left( 1 + \varepsilon \right)^{1/4} \\*[1mm]
0 & \left( 1 + \varepsilon \right)^{1/4} &
\sqrt{1 + \varepsilon} - 1
\end{array} \right).
\ee
The diagonalization of the real matrices $M_n$ and $M_i$ proceeds as
$O_\mathrm{n}^T M_\mathrm{n} O_\mathrm{n} = \sqrt{\Delta}\
\diag \left( 0,\ - \sqrt{\varepsilon},\ 1 \right)$
and $O_\mathrm{i}^T M_\mathrm{i} O_\mathrm{i} = \sqrt{\Delta}\
\diag \left( -1,\ \sqrt{1 + \varepsilon},\ 0 \right)$,
with\footnote{The remarkable and desirable properties
of mass matrices like $M_\mathrm{n}$ and $M_\mathrm{i}$
have been noticed long ago~\cite{weinberg}.}
\bs
\label{oni}
\ba
O_\mathrm{n} &=& \left( \begin{array}{ccc} 1 & 0 & 0 \\*[1mm]
0 &
\frac{1}{\sqrt{1 + \sqrt{\varepsilon}}} &
\frac{\varepsilon^{1/4}}{\sqrt{1 + \sqrt{\varepsilon}}} \\*[3mm]
0 &
- \frac{\varepsilon^{1/4}}{\sqrt{1 + \sqrt{\varepsilon}}} &
\frac{1}{\sqrt{1 + \sqrt{\varepsilon}}}
\end{array} \right),
\label{ivopd}
\\*[3mm]
O_\mathrm{i} &=& \left( \begin{array}{ccc} 0 & 0 & 1 \\*[1mm]
\frac{\left( 1 + \varepsilon \right)^{1/4}}
{\sqrt{1 + \sqrt{1 + \varepsilon}}} &
\frac{1}
{\sqrt{1 + \sqrt{1 + \varepsilon}}} &
0 \\*[3mm]
- \frac{1}
{\sqrt{1 + \sqrt{1 + \varepsilon}}} &
\frac{\left( 1 + \varepsilon \right)^{1/4}}
{\sqrt{1 + \sqrt{1 + \varepsilon}}} &
0
\end{array} \right),
\label{bvkor}
\ea
\es
We see that the mixing angle $\theta_\mathrm{b}$ appears in $O_\mathrm{b}$.
If b is n,
then
\be
\cos{\theta_\mathrm{n}} = \sqrt{\frac{1}{1 + \sqrt{\varepsilon}}}
\equiv c_\mathrm{n},
\quad
\sin{\theta_\mathrm{n}} = \sqrt{\frac{\sqrt{\varepsilon}}
{1 + \sqrt{\varepsilon}}} \equiv s_\mathrm{n}.
\ee
If b is i,
then
\be
\cos{\theta_\mathrm{i}} = \sqrt{\frac{1}{1 + \sqrt{1 + \varepsilon}}}
\equiv c_\mathrm{i},
\quad
\sin{\theta_\mathrm{i}} = \sqrt{\frac{\sqrt{1 + \varepsilon}}
{1 + \sqrt{1 + \varepsilon}}}
\equiv s_\mathrm{i},
\ee
Since $\varepsilon \sim 1 / 30$ is small,
$\theta_\mathrm{n} \sim 20^\circ$ is smallish.
On the other hand,
$\theta_\mathrm{i}$ is very close to 45 degrees,
\textit{viz.}\ almost maximal.

It turns out that,
because the mixing angle $\theta_\mathrm{n}$ is so small,
case n is not much different from the one,
treated in eqs.~(\ref{xi1})--(\ref{xi3}),
in which lepton mixing originates fully in $M_\ell$.
Because of this,
\emph{a normal neutrino mass spectrum does not work with (5, 2) textures}.

For case i,
one may write down the six possible forms
of the lepton mixing matrices.
They are
\bs
\label{evips}
\ba
\frac{M_\mathrm{i}}{\sqrt{\Delta}} = \left( \begin{array}{ccc}
0 & 0 & 0 \\ 0 & 0 & \left( 1 + \varepsilon \right)^{1/4} \\*[1mm]
0 & \left( 1 + \varepsilon \right)^{1/4} & \sqrt{1 + \varepsilon} - 1
\end{array} \right)
& \Rightarrow &
U = O_\ell^T \left( \begin{array}{ccc}
0 & 0 & 1 \\
s_\mathrm{i} e^{i \aleph} & c_\mathrm{i} e^{i \aleph} & 0 \\
- c_\mathrm{i} & s_\mathrm{i} & 0
\end{array} \right),
\no & &
\label{al}
\\
\frac{M_\mathrm{i}}{\sqrt{\Delta}} = \left( \begin{array}{ccc}
0 & \left( 1 + \varepsilon \right)^{1/4} & 0 \\*[1mm]
\left( 1 + \varepsilon \right)^{1/4} & \sqrt{1 + \varepsilon} - 1 & 0 \\
0 & 0 & 0
\end{array} \right)
& \Rightarrow &
U = O_\ell^T \left( \begin{array}{ccc}
s_\mathrm{i} e^{i \aleph} & c_\mathrm{i} e^{i \aleph} & 0 \\
- c_\mathrm{i} & s_\mathrm{i} & 0 \\ 0 & 0 & 1
\end{array} \right),
\no & &
\label{bl}
\\
\frac{M_\mathrm{i}}{\sqrt{\Delta}} = \left( \begin{array}{ccc}
\sqrt{1 + \varepsilon} - 1 & 0 & \left( 1 + \varepsilon \right)^{1/4} \\
0 & 0 & 0 \\
\left( 1 + \varepsilon \right)^{1/4} & 0 & 0
\end{array} \right)
& \Rightarrow &
U = O_\ell^T \left( \begin{array}{ccc}
- c_\mathrm{i} & s_\mathrm{i} & 0 \\ 0 & 0 & 1 \\
s_\mathrm{i} e^{i \aleph} & c_\mathrm{i} e^{i \aleph} & 0
\end{array} \right),
\no & &
\label{c}
\\
\frac{M_\mathrm{i}}{\sqrt{\Delta}} = \left( \begin{array}{ccc}
0 & 0 & 0 \\*[1mm]
0 & \sqrt{1 + \varepsilon} - 1 & \left( 1 + \varepsilon \right)^{1/4} \\
0 & \left( 1 + \varepsilon \right)^{1/4} & 0
\end{array} \right)
& \Rightarrow &
U = O_\ell^T \left( \begin{array}{ccc}
0 & 0 & 1 \\
- c_\mathrm{i} & s_\mathrm{i} & 0 \\
s_\mathrm{i} e^{i \aleph} & c_\mathrm{i} e^{i \aleph} & 0
\end{array} \right),
\no & &
\label{d}
\\
\frac{M_\mathrm{i}}{\sqrt{\Delta}} = \left( \begin{array}{ccc}
0 & 0 & \left( 1 + \varepsilon \right)^{1/4} \\
0 & 0 & 0 \\*[1mm]
\left( 1 + \varepsilon \right)^{1/4} & 0 & \sqrt{1 + \varepsilon} - 1
\end{array} \right)
& \Rightarrow &
U = O_\ell^T \left( \begin{array}{ccc}
s_\mathrm{i} e^{i \aleph} & c_\mathrm{i} e^{i \aleph} & 0 \\
0 & 0 & 1 \\
- c_\mathrm{i} & s_\mathrm{i} & 0 \\
\end{array} \right),
\no & &
\label{e}
\\
\frac{M_\mathrm{i}}{\sqrt{\Delta}} = \left( \begin{array}{ccc}
\sqrt{1 + \varepsilon} - 1 & \left( 1 + \varepsilon \right)^{1/4} & 0 \\
\left( 1 + \varepsilon \right)^{1/4} & 0 & 0 \\
0 & 0 & 0 \\*[1mm]
\end{array} \right)
& \Rightarrow &
U = O_\ell^T \left( \begin{array}{ccc}
- c_\mathrm{i} & s_\mathrm{i} & 0 \\
s_\mathrm{i} e^{i \aleph} & c_\mathrm{i} e^{i \aleph} & 0 \\
0 & 0 & 1
\end{array} \right).
\no & &
\label{f}
\ea
\es
In the forms for $\pmns$ in eqs.~(\ref{evips}),
the matrix $O_\ell$ is the real orthogonal matrix that diagonalizes $H$
according to eq.~(\ref{nbkpd}).
The matrix $O_\ell$ contains two degrees of freedom
because $H$ satisfies either $H_{12} = 0$
or $\left( H^{-1} \right)_{12} = 0$.
The matrix $\pmns$ depends on three degrees of freedom:
one of them is the phase $\aleph$ and the other two are contained in $O_\ell$.
So,
there is one non-trivial constraint on $\pmns$.

For the mass term responsible for neutrinoless double-beta decay
one finds the formula
\be
\label{bb}
\frac{m_{\beta \beta}}{\sqrt{\Delta}} = \left| \left( O_\ell \right)_{i1} \right|
\left|
\left( \sqrt{1 + \varepsilon} - 1 \right) \left( O_\ell \right)_{i1}
+ 2 \left( 1 + \varepsilon \right)^{1/4} e^{i \aleph} \left( O_\ell \right)_{j1}
\right|,
\ee
where the values for the indices $i$ and $j$ are given in table~1.
\begin{table}
\begin{center}
\renewcommand{\arraystretch}{1.2}
\begin{tabular}{|c||c|c|c|c|c|c|}
\hline
Equation for $\pmns$ & (\ref{al}) & (\ref{bl}) & (\ref{c}) &
(\ref{d}) & (\ref{e}) & (\ref{f}) \\
\hline \hline
$i$ & 3 & 2 & 1 & 2 & 3 & 1 \\
\hline
$j$ & 2 & 1 & 3 & 3 & 1 & 2 \\
\hline
\end{tabular}
\end{center}
\caption{The indices $i$ and $j$ to be used in eq.~(\ref{bb}).
\label{table1}}
\end{table}

One should note that eqs.~(\ref{d})--(\ref{f})
correspond to eqs.~(\ref{al})--(\ref{c}),
respectively,
after an interchange between $\nu_\mu$ and $\nu_\tau$.
This interchange is equivalent,
in the standard parametrization of $\pmns$,
to the transformations $\cos{\theta_{23}} \leftrightarrow \sin{\theta_{23}}$
and $\cos{\delta} \rightarrow - \cos{\delta}$.
In so far as the extant phenomenological data
are approximately invariant under
$\cos{\theta_{23}} \leftrightarrow \sin{\theta_{23}}$,
one may anticipate that the predictions of eqs.~(\ref{d})--(\ref{f})
for $\cos{\delta}$ will be approximately symmetric
to the corresponding predictions of eqs.~(\ref{al})--(\ref{c}).

\begin{figure}[hbt]
\centering
\includegraphics[width=10cm,angle=0]{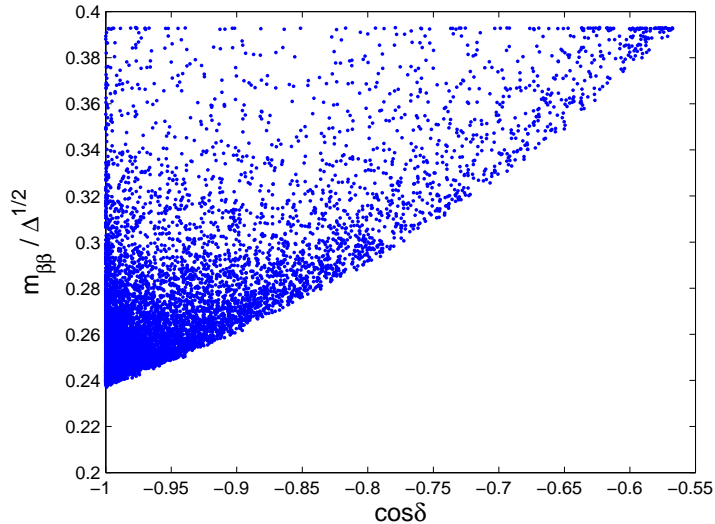}
\caption{$m_{\beta \beta} \left/ \sqrt{\Delta}\right.$ \textit{vs} $\cos{\delta}$
in a numeric scan for models with (5, 2) textures.}
\label{fig:mbbc}
\end{figure}

We have found numerically that
all six eqs.~(\ref{evips}) are able to fit $\pmns$
provided $H_{12} = 0$,
but they are unable to achieve that fit when
$\left( H^{-1} \right)_{12} = 0$.
Furthermore,
the predictions of eqs.~(\ref{al})--(\ref{c})
(with $H_{12} = 0$)
are all very similar
(but not really identical)
among themselves.
In all those cases,
one must have a rather large solar mixing angle,
$\sin^2{\theta_{12}} \gtrsim 0.3$.
The prediction of eqs.~(\ref{al})--(\ref{c})
for the Dirac phase is $\cos{\delta} \lesssim -0.6$,
while eqs.~(\ref{d})--(\ref{f}) make the symmetric prediction
$\cos{\delta} \gtrsim 0.6$.
The prediction for neutrinoless double-beta decay is
$0.24 \lesssim m_{\beta \beta} \left/ \sqrt{\Delta} \right. \lesssim 0.4$.
We can see these predictions displayed in fig.~\ref{fig:mbbc},
in which we plot $m_{\beta \beta} \left/ \sqrt{\Delta} \right.$
against $\cos{\delta}$.
Each point in the plot corresponds to some definite values
for the parameters of the model---neutrino oscillation observables,
phase $\aleph$,
and two entries of the matrix $H$).
For definiteness,
these predictions are based on the use of the phenomenological $3 \sigma$
data in ref.~\cite{Forero:2014bxa};
other phenomenological fits to the data---see refs.~\cite{Fogli:2012ua}
and~\cite{GonzalezGarcia:2012sz}---can hardly yield
much too different results.

\section{(4, 3) textures}

Since there are no massless charged leptons,
the determinant of $M_\ell$ must be nonzero.
Therefore,
after an adequate reordering of the rows and columns of $M_\ell$,
\be
\label{mell}
M_\ell = \left( \begin{array}{ccc}
t_1 & 0 & 0 \\ 0 & t_2 & 0 \\ 0 & t_3 & t_4
\end{array} \right).
\ee
Therefore,
\be
\label{Hell}
H = \left( \begin{array}{ccc}
\left| t_1 \right|^2 & 0 & 0 \\
0 & \left| t_2 \right|^2 & \left| t_2 t_3 \right| e^{i \gamma} \\
0 & \left| t_2 t_3 \right| e^{- i \gamma} &
\left| t_3 \right|^2 + \left| t_4 \right|^2
\end{array} \right),
\ee
where $\gamma \equiv \arg{\left( t_2 t_3^\ast \right)}$.
From eq.~(\ref{bkpdl}),
the columns of $U_L$ are the normalized eigenvectors of $H$.
It is clear from eq.~(\ref{Hell}) that
one of the eigenvalues of $H$ is $\left| t_1 \right|^2$
and the corresponding normalized eigenvector is $\left( 1,\ 0,\ 0 \right)^T$.
Therefore,
either
\bs
\label{UL}
\ba
\label{UL1}
U_L &=& \left( \begin{array}{ccc}
1 & 0 & 0 \\ 0 & \cos{\theta} & e^{i \gamma} \sin{\theta} \\
0 & - e^{- i \gamma} \sin{\theta} & \cos{\theta}
\end{array} \right) X, \quad \mathrm{or}
\\
\label{UL2}
U_L &=& \left( \begin{array}{ccc}
0 & 1 & 0 \\
\cos{\theta} & 0 & e^{i \gamma} \sin{\theta} \\
- e^{- i \gamma} \sin{\theta} & 0 & \cos{\theta}
\end{array} \right) X, \quad \mathrm{or}
\\
\label{UL3}
U_L &=& \left( \begin{array}{ccc}
0 & 0 & 1 \\
\cos{\theta} & e^{i \gamma} \sin{\theta} & 0 \\
- e^{- i \gamma} \sin{\theta} & \cos{\theta} & 0
\end{array} \right) X,
\ea
\es
where $X$ is a diagonal unitary matrix
containing the phases of the eigenvectors of $H$;
those phases are meaningless.
Equation~(\ref{UL1}) holds if $\left| t_1 \right| = m_e$,
eq.~(\ref{UL2}) holds if $\left| t_1 \right| = m_\mu$,
and eq.~(\ref{UL3}) holds if $\left| t_1 \right| = m_\tau$.
The angle $\theta$ is fixed by
\be
\tan{2 \theta} = \frac{2 \left| t_2 t_3 \right|}
{\left| t_3 \right|^2 + \left| t_4 \right|^2 - \left| t_2 \right|^2}.
\ee

We assume that only three out of the six independent matrix elements
of $M_\nu$ are nonzero.
Therefore there are $\left. \left( 6 \times 5 \times 4 \right) \right/ 3! = 20$
possible forms for $M_\nu$. They are
\ba
\label{ubiop}
M_\nu &\sim& \left( \begin{array}{ccc}
\times & 0 & 0 \\ 0 & \times & 0 \\ 0 & 0 & \times
\end{array} \right),
\\
M_\nu &\sim& \left( \begin{array}{ccc}
\times & \times & 0 \\ \times & \times & 0 \\ 0 & 0 & 0
\end{array} \right),
\\
M_\nu &\sim& \left( \begin{array}{ccc}
\times & 0 & \times \\ 0 & \times & 0 \\ \times & 0 & 0
\end{array} \right),
\\
M_\nu &\sim& \left( \begin{array}{ccc}
\times & 0 & 0 \\ 0 & \times & \times \\ 0 & \times & 0
\end{array} \right),
\\
M_\nu &\sim& \left( \begin{array}{ccc}
\times & \times & 0 \\ \times & 0 & 0 \\ 0 & 0 & \times
\end{array} \right),
\\
M_\nu &\sim& \left( \begin{array}{ccc}
\times & 0 & \times \\ 0 & 0 & 0 \\ \times & 0 & \times
\end{array} \right),
\\
M_\nu &\sim& \left( \begin{array}{ccc}
\times & 0 & 0 \\ 0 & 0 & \times \\ 0 & \times & \times
\end{array} \right),
\\
M_\nu &\sim& \left( \begin{array}{ccc}
0 & \times & 0 \\ \times & \times & 0 \\ 0 & 0 & \times
\end{array} \right),
\\
M_\nu &\sim& \left( \begin{array}{ccc}
0 & 0 & \times \\ 0 & \times & 0 \\ \times & 0 & \times
\end{array} \right),
\\
M_\nu &\sim& \left( \begin{array}{ccc}
0 & 0 & 0 \\ 0 & \times & \times \\ 0 & \times & \times
\end{array} \right),
\ea
\ba
\label{a}
M_\nu &\sim& \left( \begin{array}{ccc}
\times & \times & \times \\ \times & 0 & 0 \\ \times & 0 & 0
\end{array} \right),
\label{11} \\
\label{b}
M_\nu &\sim& \left( \begin{array}{ccc}
\times & \times & 0 \\ \times & 0 & \times \\ 0 & \times & 0
\end{array} \right),
\\
M_\nu &\sim& \left( \begin{array}{ccc}
\times & 0 & \times \\ 0 & 0 & \times \\ \times & \times & 0
\end{array} \right),
\\
M_\nu &\sim& \left( \begin{array}{ccc}
0 & \times & \times \\ \times & \times & 0 \\ \times & 0 & 0
\end{array} \right),
\\
M_\nu &\sim& \left( \begin{array}{ccc}
0 & \times & 0 \\ \times & \times & \times \\ 0 & \times & 0
\end{array} \right),
\\
M_\nu &\sim& \left( \begin{array}{ccc}
0 & 0 & \times \\ 0 & \times & \times \\ \times & \times & 0
\end{array} \right),
\\
M_\nu &\sim& \left( \begin{array}{ccc}
0 & \times & \times \\ \times & 0 & 0 \\ \times & 0 & \times
\end{array} \right),
\\
M_\nu &\sim& \left( \begin{array}{ccc}
0 & \times & 0 \\ \times & 0 & \times \\ 0 & \times & \times
\end{array} \right),
\\
M_\nu &\sim& \left( \begin{array}{ccc}
0 & 0 & \times \\ 0 & 0 & \times \\ \times & \times & \times
\end{array} \right),
\\
M_\nu &\sim& \left( \begin{array}{ccc}
0 & \times & \times \\ \times & 0 & \times \\ \times & \times & 0
\end{array} \right).
\label{20}
\ea
%

Let $Z$ be any of the six matrices in the group $S_3$ of eq.~(\ref{ss33}).
Those matrices are orthogonal,
hence $Z^{-1} = Z^T$.
Interchanging the rows and columns of $M_\nu$
is equivalent to making $M_\nu \to Z M_\nu Z^T$.
But $U_\nu \to Z U_\nu$ when this happens.
Therefore $\pmns \to U_L^\dagger Z U_\nu$.
This is equivalent to letting $U_L \to Z^\dagger U_L$,
which corresponds to a reordering of the rows of $U_L$.
We conclude that,
provided one allows for a reordering of the rows
of the three possibilities for $U_L$ in eqs.~(\ref{UL}),
one is free to avoid considering separately two matrices $M_\nu$
which differ only by an interchange of their rows and columns.
In this way,
out of the 20 forms for $M_\nu$ in eqs.~(\ref{ubiop}--\ref{20}),
one only needs to consider the following six:
\bs
\ba
\label{d1}
M_\nu &\sim& \left( \begin{array}{ccc}
\times & 0 & 0 \\ 0 & \times & 0 \\ 0 & 0 & \times
\end{array} \right),
\\
\label{d2}
M_\nu &\sim& \left( \begin{array}{ccc}
\times & 0 & \times \\ 0 & 0 & 0 \\ \times & 0 &  \times
\end{array} \right),
\\
\label{d3}
M_\nu &\sim& \left( \begin{array}{ccc}
\times & \times & 0 \\ \times & 0 & 0 \\ 0 & 0 & \times
\end{array} \right),
\\
\label{d4}
M_\nu &\sim& \left( \begin{array}{ccc}
\times & \times & \times \\ \times & 0 & 0 \\ \times & 0 & 0
\end{array} \right),
\\
\label{d5}
M_\nu &\sim& \left( \begin{array}{ccc}
\times & \times & 0 \\ \times & 0 & \times \\ 0 & \times & 0
\end{array} \right),
\\
\label{d6}
M_\nu &\sim& \left( \begin{array}{ccc}
0 & \times & \times \\ \times & 0 & \times \\ \times & \times & 0
\end{array} \right).
\ea
\es
The first three of these forms for $M_\nu$
are excluded when taken in conjunction with the $U_L$ matrices
in eqs.~(\ref{UL}).
Indeed,
eq.~(\ref{d1}) leads to $\pmns$ with four zero matrix elements;
either eq.~(\ref{d2}) or eq.~(\ref{d3}) lead to $\pmns$
with one zero matrix element;
and both those situations are phenomenologically excluded.
The only viable forms of $M_\nu$ are those that give rise
to genuine $3 \times 3$ mixing in $M_\nu$,
\textit{viz.}\ eqs.~(\ref{d4}--\ref{d6}).

One may,
without lack of generality,
assume the three nonzero matrix elements of $M_\nu$ to be real and positive,
because,
for each of the three matrices in eqs.~(\ref{d4}--\ref{d6}),
there is a diagonal unitary matrix
$Y_\psi = \diag \left( e^{i \psi_1},\ e^{i \psi_2},\ e^{i \psi_3} \right)$
such that $Y_\psi M_\nu Y_\psi$ is real and has positive nonzero matrix elements.
We may thus write the matrices
\be
\label{mk}
M_A = \left( \begin{array}{ccc}
a & d & b \\ d & 0 & 0 \\ b & 0 & 0
\end{array} \right),
\quad
M_B = \left( \begin{array}{ccc}
a & b & 0 \\ b & 0 & d \\ 0 & d & 0
\end{array} \right),
\quad
M_C = \left( \begin{array}{ccc}
0 & a & b \\ a & 0 & d \\ b & d & 0
\end{array} \right),
\ee
where $a$,
$b$,
and $d$ are positive.
One has $M_\nu = Y_\psi^\ast M_K Y_\psi^\ast$,
where $K$ may be either $A$,
$B$,
or $C$.

The matrix $M_K$ is diagonalized by the orthogonal matrix $O_K$:
\be
\label{ok}
O_K^T M_K O_K = \diag \left( \mu_1,\ \mu_2,\ \mu_3 \right).
\ee
The real numbers $\mu_k$ ($k = 1, 2, 3$) are the eigenvalues of $M_K$;
$\left| \mu_k \right| = m_k$ are the neutrino masses.

From eq.~(\ref{Unu}),
\be
U_\nu = Y_\psi O_K Y^\prime;
\ee
the matrix $Y^\prime$ is a diagonal unitary matrix
which affects the transformation $\mu_k \to m_k$
in the following way:
$Y^\prime_{kk} = 1$ if $\mu_k > 0$ and $Y^\prime_{kk} = i$ if $\mu_k < 0$.
So,
from eq.~(\ref{pmns}),
$\pmns = U_L^\dagger Y_\psi O_K Y^\prime$,
where $U_L$ is either one of the matrices in eqs.~(\ref{UL})
or one of them with the rows interchanged.

The matrix $U_L^\dagger Y_\psi$ contains four phases---one phase $\gamma$
in $U_L^\dagger$ and three phases $\psi_{1,2,3}$ in $Y_\psi$.
One may pull
three of those phases to the left-hand side of $U_L^\dagger$,
leaving at its right-hand side only one phase---let $\chi$ denote it.
Suppose for instance that eq.~(\ref{UL1}) holds,
then
\ba
U_L^\dagger Y_\psi &=&
X^\ast \left( \begin{array}{ccc}
e^{i \psi_1} & 0 & 0 \\
0 & e^{i \psi_2} \cos{\theta} & - e^{i \left( \psi_3 + \gamma \right)} \sin{\theta} \\
0 & e^{i \left( \psi_2 -  \gamma \right)} \sin{\theta} & e^{i \psi_3} \cos{\theta}
\end{array} \right)
\no &=&
X^\ast\ \mathrm{diag} \left( e^{i \psi_1},\ e^{i \psi_2},\
e^{i \left( \psi_2 - \gamma \right)} \right)
\left( \begin{array}{ccc}
1 & 0 & 0 \\
0 & \cos{\theta} & - e^{i \chi} \sin{\theta} \\
0 & \sin{\theta} & e^{i \chi} \cos{\theta}
\end{array} \right)
\no &=&
X^\prime
\left( \begin{array}{ccc}
1 & 0 & 0 \\
0 & \cos{\theta} & - e^{i \chi} \sin{\theta} \\
0 & \sin{\theta} & e^{i \chi} \cos{\theta}
\end{array} \right),
\label{pmnsform}
\ea
where $\chi \equiv \psi_3 - \psi_2 + \gamma$.
The matrix $X^\prime \equiv X^\ast\ \diag \left( e^{i \psi_1},\ e^{i \psi_2},\
e^{i \left( \psi_2 - \gamma \right)} \right)$ contains unphysical phases.

Thus,
there are 18 possible forms for $\pmns$ in $(4, 3)$ textures.
Let $Kp$ denote those 18 forms,
where $K$ may be either $A$,
$B$,
or $C$.
If $K = B$,
then the number $p$ may be $1, 2, \ldots, 9$:
\bs
\label{ABC}
\ba
\label{1}
\pmns &=& \left( \begin{array}{ccc}
1 & 0 & 0 \\
0 & \cos{\theta} & - e^{i \chi} \sin{\theta} \\
0 & \sin{\theta} & e^{i \chi} \cos{\theta}
\end{array} \right) O_B Y^\prime
\quad (\mathrm{form}\ B1);
\\
\label{b2}
\pmns &=& \left( \begin{array}{ccc}
0 & \cos{\theta} & - e^{i \chi} \sin{\theta} \\
1 & 0 & 0 \\
0 & \sin{\theta} & e^{i \chi} \cos{\theta}
\end{array} \right) O_B Y^\prime
\quad (\mathrm{form}\ B2);
\\
\label{b3}
\pmns &=& \left( \begin{array}{ccc}
0 & \cos{\theta} & - e^{i \chi} \sin{\theta} \\
0 & \sin{\theta} & e^{i \chi} \cos{\theta} \\
1 & 0 & 0
\end{array} \right) O_B Y^\prime
\quad (\mathrm{form}\ B3);
\\
\pmns &=& \left( \begin{array}{ccc}
0 & 1 & 0 \\
\cos{\theta} & 0 & - e^{i \chi} \sin{\theta} \\
\sin{\theta} & 0 & e^{i \chi} \cos{\theta}
\end{array} \right) O_B Y^\prime
\quad (\mathrm{form}\ B4);
\\
\pmns &=& \left( \begin{array}{ccc}
\cos{\theta} & 0 & - e^{i \chi} \sin{\theta} \\
0 & 1 & 0 \\
\sin{\theta} & 0 & e^{i \chi} \cos{\theta}
\end{array} \right) O_B Y^\prime
\quad (\mathrm{form}\ B5);
\\
\pmns &=& \left( \begin{array}{ccc}
\cos{\theta} & 0 & - e^{i \chi} \sin{\theta} \\
\sin{\theta} & 0 & e^{i \chi} \cos{\theta} \\
0 & 1 & 0
\end{array} \right) O_B Y^\prime
\quad (\mathrm{form}\ B6);
\\
\pmns &=& \left( \begin{array}{ccc}
0 & 0 & 1 \\
\cos{\theta} & - e^{i \chi} \sin{\theta} & 0 \\
\sin{\theta} & e^{i \chi} \cos{\theta} & 0
\end{array} \right) O_B Y^\prime
\quad (\mathrm{form}\ B7);
\\
\label{b8}
\pmns &=& \left( \begin{array}{ccc}
\cos{\theta} & - e^{i \chi} \sin{\theta} & 0 \\
0 & 0 & 1 \\
\sin{\theta} & e^{i \chi} \cos{\theta} & 0
\end{array} \right) O_B Y^\prime
\quad (\mathrm{form}\ B8);
\\
\label{b9}
\pmns &=& \left( \begin{array}{ccc}
\cos{\theta} & - e^{i \chi} \sin{\theta} & 0 \\
\sin{\theta} & e^{i \chi} \cos{\theta} & 0 \\
0 & 0 & 1
\end{array} \right) O_B Y^\prime
\quad (\mathrm{form}\ B9).
\ea
\es
The real orthogonal matrix $O_B$
diagonalizes the real symmetric matrix $M_B$,
see eqs.~(\ref{mk}) and~(\ref{ok}).

When one interchanges the second and third rows and columns
in the matrix $M_A$ one obtains the same matrix with $b$ and $d$ interchanged;
this is just a meaningless renaming of parameters.
Similarly,
any permutation of the rows and columns of $M_C$ is equivalent
to a renaming of the parameters $a$,
$b$,
and $d$.
Therefore,
there are nine more possible forms for $\pmns$:
\bs
\label{ABC2}
\ba
\label{10}
\pmns &=& \left( \begin{array}{ccc}
1 & 0 & 0 \\
0 & \cos{\theta} & - e^{i \chi} \sin{\theta} \\
0 & \sin{\theta} & e^{i \chi} \cos{\theta}
\end{array} \right) O_A Y^\prime
\quad (\mathrm{form}\ A1);
\\
\pmns &=& \left( \begin{array}{ccc}
0 & \cos{\theta} & - e^{i \chi} \sin{\theta} \\
1 & 0 & 0 \\
0 & \sin{\theta} & e^{i \chi} \cos{\theta}
\end{array} \right) O_A Y^\prime
\quad (\mathrm{form}\ A2);
\\
\pmns &=& \left( \begin{array}{ccc}
0 & \cos{\theta} & - e^{i \chi} \sin{\theta} \\
0 & \sin{\theta} & e^{i \chi} \cos{\theta} \\
1 & 0 & 0
\end{array} \right) O_A Y^\prime
\quad (\mathrm{form}\ A3);
\\
\pmns &=& \left( \begin{array}{ccc}
0 & 1 & 0 \\
\cos{\theta} & 0 & - e^{i \chi} \sin{\theta} \\
\sin{\theta} & 0 & e^{i \chi} \cos{\theta}
\end{array} \right) O_A Y^\prime
\quad (\mathrm{form}\ A4);
\\
\label{155}
\pmns &=& \left( \begin{array}{ccc}
\cos{\theta} & 0 & - e^{i \chi} \sin{\theta} \\
0 & 1 & 0 \\
\sin{\theta} & 0 & e^{i \chi} \cos{\theta}
\end{array} \right) O_A Y^\prime
\quad (\mathrm{form}\ A5);
\\
\label{16}
\pmns &=& \left( \begin{array}{ccc}
\cos{\theta} & 0 & - e^{i \chi} \sin{\theta} \\
\sin{\theta} & 0 & e^{i \chi} \cos{\theta} \\
0 & 1 & 0
\end{array} \right) O_A Y^\prime
\quad (\mathrm{form}\ A6);
\\
\label{70}
\pmns &=& \left( \begin{array}{ccc}
1 & 0 & 0 \\
0 & \cos{\theta} & - e^{i \chi} \sin{\theta} \\
0 & \sin{\theta} & e^{i \chi} \cos{\theta}
\end{array} \right) O_C Y^\prime
\quad (\mathrm{form}\ C1);
\\
\label{80}
\pmns &=& \left( \begin{array}{ccc}
0 & \cos{\theta} & - e^{i \chi} \sin{\theta} \\
1 & 0 & 0 \\
0 & \sin{\theta} & e^{i \chi} \cos{\theta}
\end{array} \right) O_C Y^\prime
\quad (\mathrm{form}\ C2);
\\
\label{90}
\pmns &=& \left( \begin{array}{ccc}
0 & \cos{\theta} & - e^{i \chi} \sin{\theta} \\
0 & \sin{\theta} & e^{i \chi} \cos{\theta} \\
1 & 0 & 0
\end{array} \right) O_C Y^\prime
\quad (\mathrm{form}\ C3).
\ea
\es

In eqs.~(\ref{ABC}) and~(\ref{ABC2})
the angle $\theta$ and the phase $\chi$ are free parameters,
to be adjusted in order to obtain a good fit of $\pmns$.
The diagonal unitary matrix $Y^\prime$ is in practice
irrelevant for phenomenology.

For the parameter of neutrinoless double-beta decay $m_{\beta\beta}$
one easily derives the formulae
\bs
\ba
m_{\beta\beta} &=& 0 \quad \mathrm{for\ forms}\ A2,\ A3,\ A4,\ B4,\ B7,\
\mathrm{and}\ C1; \\
m_{\beta\beta} &=& a \quad \mathrm{for\ forms}\ A1\
\mathrm{and}\ B1; \\
m_{\beta\beta} &=& a \cos^2{\theta} \quad \mathrm{for\ forms}\ B5\
\mathrm{and}\ B6; \\
m_{\beta\beta} &=& d \sin{2 \theta} \quad \mathrm{for\ forms}\ B2,\ B3,\ C2,\
\mathrm{and}\ C3; \\
m_{\beta\beta} &=& \left| a \cos^2{\theta}
- b e^{- i \chi} \sin{2 \theta} \right|  \quad \mathrm{for\ forms}\ A5,\ A6,\ B8,\
\mathrm{and}\ B9. \hspace*{7mm}
\ea
\es

\subsection{Forms $A1$--$6$}

We first consider the matrix $M_A$ in the first eq.~(\ref{mk})
and its diagonalizing matrix $O_A$.
It is convenient to define $f \equiv \sqrt{b^2 + d^2}$
and the angle $\varphi$:
\be
\cos{\varphi} = \frac{b}{f}, \quad \sin{\varphi} = \frac{d}{f}.
\ee
The matrix $M_A$ has vanishing determinant.
Therefore,
\emph{one neutrino is massless}\/ and eqs.~(\ref{jbipd}--\ref{vkpst}) apply.
In case n,
\bs
\ba
M_A &=& \sqrt{\Delta} \left( \begin{array}{ccc}
1 - \sqrt{\varepsilon} &
\varepsilon^{1/4} \sin{\varphi} &
\varepsilon^{1/4} \cos{\varphi} \\*[1mm]
\varepsilon^{1/4} \sin{\varphi} & 0 & 0 \\*[1mm]
\varepsilon^{1/4} \cos{\varphi} & 0 & 0
\end{array} \right),
\\*[3mm]
\label{O1normal}
O_A &=& \left( \begin{array}{ccc}
0 &
\frac{\varepsilon^{1/4}}{\sqrt{1 + \sqrt{\varepsilon}}} &
\frac{1}{\sqrt{1 + \sqrt{\varepsilon}}} \\*[3mm]
\cos{\varphi} &
- \frac{1}{\sqrt{1 + \sqrt{\varepsilon}}}\, \sin{\varphi} &
\frac{\varepsilon^{1/4}}{\sqrt{1 + \sqrt{\varepsilon}}}\, \sin{\varphi} \\*[3mm]
- \sin{\varphi} &
- \frac{1}{\sqrt{1 + \sqrt{\varepsilon}}}\, \cos{\varphi} &
\frac{\varepsilon^{1/4}}{\sqrt{1 + \sqrt{\varepsilon}}}\, \cos{\varphi}
\end{array} \right).
\ea
\es
In case i,
\bs
\ba
M_A &=& \sqrt{\Delta} \left( \begin{array}{ccc}
\sqrt{1 + \varepsilon} - 1 &
\left( 1 + \varepsilon \right)^{1/4} \sin{\varphi} &
\left( 1 + \varepsilon \right)^{1/4} \cos{\varphi} \\*[1mm]
\left( 1 + \varepsilon \right)^{1/4} \sin{\varphi} & 0 & 0 \\*[1mm]
\left( 1 + \varepsilon \right)^{1/4} \cos{\varphi} & 0 & 0
\end{array} \right), \hspace*{8mm}
\\*[3mm]
\label{O1inverted}
O_A &=& \left( \begin{array}{ccc}
\frac{1}{\sqrt{1 + \sqrt{1 + \varepsilon}}} &
\frac{\left( 1 + \varepsilon \right)^{1/4}}
{\sqrt{1 + \sqrt{1 + \varepsilon}}} &
0 \\*[3mm]
- \frac{\left( 1 + \varepsilon \right)^{1/4}}
{\sqrt{1 + \sqrt{1 + \varepsilon}}}\, \sin{\varphi} &
\frac{1}{\sqrt{1 + \sqrt{1 + \varepsilon}}}\, \sin{\varphi} &
\cos{\varphi} \\*[3mm]
- \frac{\left( 1 + \varepsilon \right)^{1/4}}
{\sqrt{1 + \sqrt{1 + \varepsilon}}}\, \cos{\varphi} &
\frac{1}{\sqrt{1 + \sqrt{1 + \varepsilon}}}\, \cos{\varphi} &
- \sin{\varphi}
\end{array} \right).
\ea
\es

It is clear from eqs.~(\ref{10}--\ref{16}) that one row of $\pmns$
must coincide,
but for the phases contained in $Y^\prime$,
with a row of $O_A$.
But,
no row of the matrix $O_A$ in eq.~(\ref{O1normal})
may possibly coincide with a row of $\pmns$,
therefore \emph{case n is excluded}.
This is because:
\begin{enumerate}
\item The first row of $O_A$ in eq.~(\ref{O1normal})
contains a zero matrix element,
while no matrix element of $\pmns$ vanishes.
\item In the second and third rows of eq.~(\ref{O1normal}),
the second entry is larger in modulus than the third entry
by a factor $\varepsilon^{-1/4} \approx 2.4$;
this factor is much too small
for what is observed in the first row of $\pmns$
and much too large
for what is observed in the second and third rows of $\pmns$.
\end{enumerate}

Coming to case i,
either the second row or the third row of $O_A$ in eq.~(\ref{O1inverted})
may coincide with either the second row or the third row of $\pmns$.
This is because those rows of $O_A$ feature
a first entry which is larger in modulus than the second entry
by a factor $\left( 1 + \varepsilon \right)^{1/4} \approx 1$;
this is compatible with what occurs in either the second or third row
of $\pmns$.
Therefore,
\emph{models $A5,6$ are viable}
(although with some deviation from the mean values of the mixing angles)
in case i.

\begin{figure}[hbt]
\centering
\includegraphics[width=10cm,angle=0]{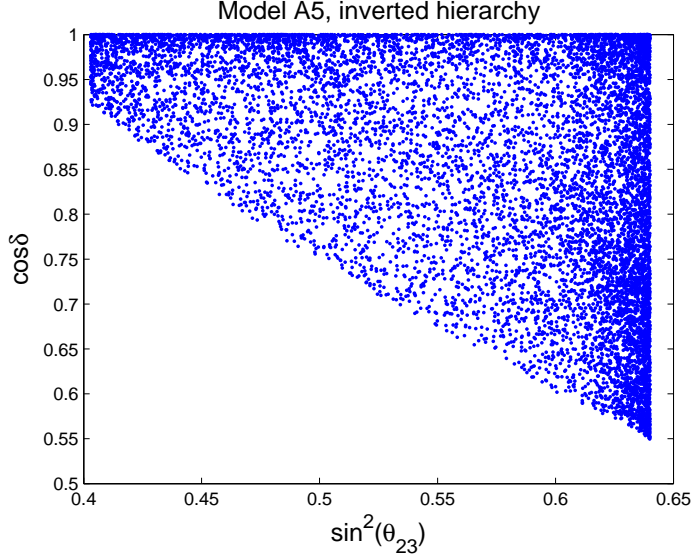}
\caption{$\cos{\delta}$ \textit{vs} $\sin^2\theta_{23}$ for form $A5$.
The numeric scan was made using the $3\, \sigma$ data
of ref.~\cite{Forero:2014bxa}.}
\label{fig:A5a}
\end{figure}

\begin{figure}[hbt]
\centering
\includegraphics[width=10cm,angle=0]{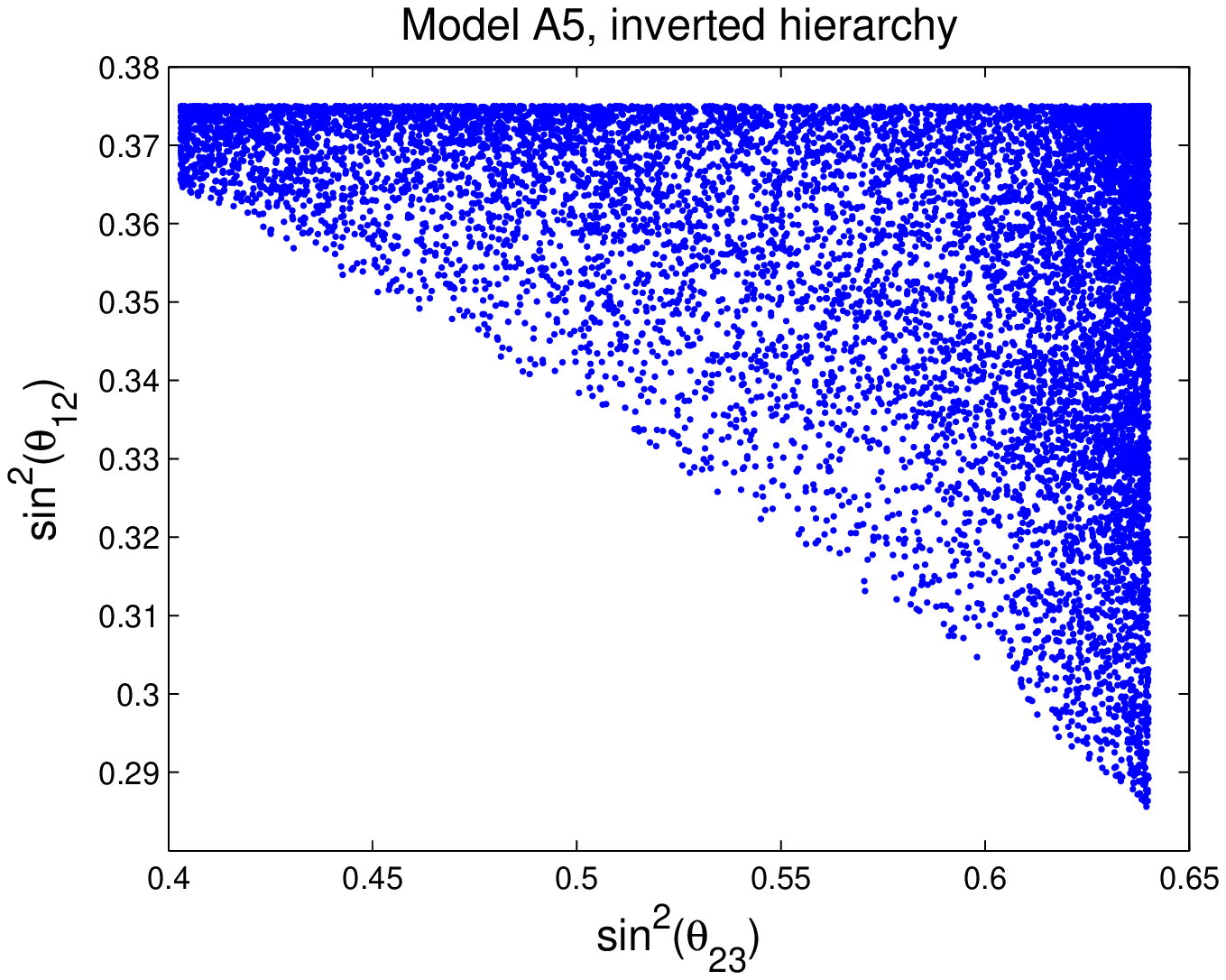}
\caption{$\sin^2\theta_{12}$ \textit{vs} $\sin^2\theta_{23}$ for form $A5$.
The numeric scan was made using the $3\, \sigma$ data
of ref.~\cite{Forero:2014bxa}.}
\label{fig:A5b}
\end{figure}

Numerically,
we have found that the form $A5$ for $\pmns$ works
(with the $3\, \sigma$ data of ref.~\cite{Forero:2014bxa})
provided $\cos{\delta} \ge 0.55$ when $\sin^2{\theta_{23}} = 0.64$;
for $\theta_{23}$ in the first octant $\cos{\delta}$ must be even closer to 1,
in particular $\cos{\delta} \ge 0.92$ for $\sin^2{\theta_{23}} = 0.40$.
The mixing angle $\theta_{12}$ must also be relatively large:
$\sin^2{\theta_{12}} > 0.285$ for $\sin^2{\theta_{23}} = 0.64$
and $\sin^2{\theta_{12}} > 0.365$ fos $\sin^2{\theta_{23}} = 0.40$.
These correlations between the angles $\theta_{12}$,
$\theta_{23}$ and the phase $\delta$ can be appreciated in figs.~\ref{fig:A5a},
\ref{fig:A5b}.
For form $A6$ of $\pmns$ the results are analogous to those of form $A5$,
except that $\cos{\delta}$ is negative instead of positive
and $\theta_{23}$ is preferred to be in the first octant
instead of in the second one.

Neutrinoless double beta decay is governed by
$0.25 < m_{\beta\beta} \left/ \sqrt{\Delta} \right. < 0.33$
in forms $A5,6$.

\subsection{Forms $C1$--$3$}

We next consider the matrix $M_C$ in the third eq.~(\ref{mk})
and its diagonalizing matrix $O_C$,
\textit{cf.}~eq.~(\ref{ok}) with $K = C$.
Since the trace of $M_C$ is zero,
$\mu_1 + \mu_2 + \mu_3 = 0$.
Also,
$\mu_1 \mu_2 \mu_3 = 2 a b d > 0$
and $\mu_1 \mu_2 + \mu_1 \mu_3 + \mu_2 \mu_3 = -a^2 - b^2 - d^2 < 0$.
Therefore,
the largest $\mu_k$ in absolute value is positive
and the other two $\mu_k$ are negative.
Thus,
in case n
\bs
\label{buopt}
\ba
\frac{\mu_1}{\sqrt{\Delta}} &=& - \sqrt{\frac{- 1 - \varepsilon
+ 2 \sqrt{1 - \varepsilon + \varepsilon^2}}{3}},
\\
\frac{\mu_2}{\sqrt{\Delta}} &=& - \sqrt{\frac{- 1 + 2 \varepsilon
+ 2 \sqrt{1 - \varepsilon + \varepsilon^2}}{3}},
\\
\frac{\mu_3}{\sqrt{\Delta}} &=& \sqrt{\frac{2 - \varepsilon
+ 2 \sqrt{1 - \varepsilon + \varepsilon^2}}{3}};
\ea
\es
while in case i
\bs
\label{lspey}
\ba
\frac{\mu_1}{\sqrt{\Delta}} &=& - \sqrt{\frac{1 - \varepsilon
+ 2 \sqrt{1 + \varepsilon + \varepsilon^2}}{3}},
\\
\frac{\mu_2}{\sqrt{\Delta}} &=& \sqrt{\frac{1 + 2 \varepsilon
+ 2 \sqrt{1 + \varepsilon + \varepsilon^2}}{3}},
\\
\frac{\mu_3}{\sqrt{\Delta}} &=& - \sqrt{\frac{- 2 - \varepsilon
+ 2 \sqrt{1 + \varepsilon + \varepsilon^2}}{3}}.
\ea
\es
Therefore,
\bs
\ba
\frac{m_1 + m_2 + m_3}{\sqrt{\Delta}} &\approx&
\frac{4 - \varepsilon}{\sqrt{3}} \quad \mathrm{in\ case\ n},
\\
\frac{m_1 + m_2 + m_3}{\sqrt{\Delta}} &\approx&
2 + \varepsilon \quad \mathrm{in\ case\ i}.
\ea
\es

\begin{figure}[hbt]
\centering
\includegraphics[width=10cm,angle=0]{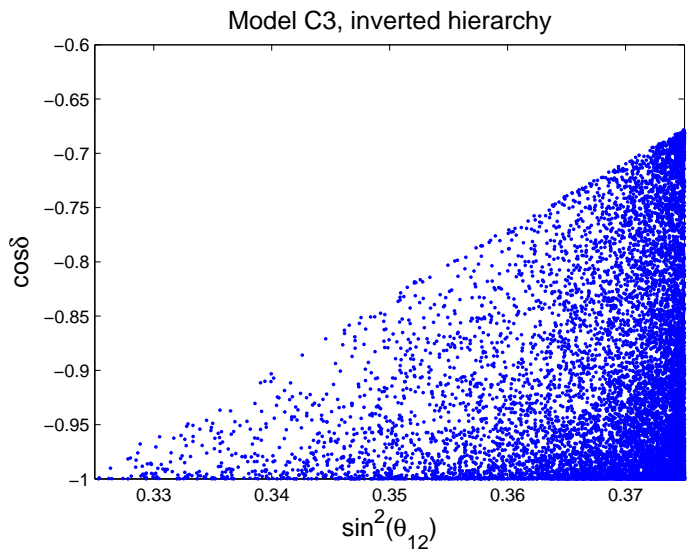}
\caption{$\cos{\delta}$ \textit{vs} $\sin^2\theta_{12}$ for form $C3$.
The numeric scan made using the $3\, \sigma$ data
of ref.~\cite{Forero:2014bxa}.}
\label{fig:C3}
\end{figure}

According to eqs.~(\ref{70}--\ref{90}),
if the PMNS matrix is of form $Ci$
($i = 1, 2, 3$)
then its $i$'th row coincides,
in the moduli of its matrix elements,
with the first row of $O_C$.
It follows from eq.~(\ref{ok}) that
\be
\left( M_C \right)_{11} = \sum_{j=1}^3
\mu_j \left[ \left( O_C \right)_{1j} \right]^2
= 0.
\label{bkpdt}
\ee
Equation~(\ref{bkpdt}),
together with the normalization of the first row of $O_C$,
yield
\bs
\label{bopdd}
\ba
\left[ \left( O_C \right)_{11} \right]^2 &=&
\frac{\mu_2 + \left( \mu_3 - \mu_2 \right)
\left[ \left( O_C \right)_{13} \right]^2}{\mu_2 - \mu_1},
\\
\left[ \left( O_C \right)_{12} \right]^2 &=&
\frac{- \mu_1 + \left( \mu_1 - \mu_3 \right)
\left[ \left( O_C \right)_{13} \right]^2}{\mu_2 - \mu_1}.
\ea
\es
Thus,
when $\pmns$ has the form $Ci$,
\be
\label{ncops}
\frac{\left| U_{i1} \right|^2}{\left| U_{i2} \right|^2}
= \frac{\mu_2 + \left( \mu_3 - \mu_2 \right)
\left| U_{i3} \right|^2}{- \mu_1 + \left( \mu_1 - \mu_3 \right)
\left| U_{i3} \right]^2}.
\ee
One may use the expressions of $\mu_{1,2,3}$
in either eqs.~(\ref{buopt}) or eqs.~(\ref{lspey})---for cases n and i,
respectively---together with $\left| U_{i3} \right|^2$
to compute $\left| U_{i1} / U_{i2} \right|^2$ through eq.~(\ref{ncops}).
One can in this way find out for which values of $\varepsilon$
and of the parameters of $U$ the form $Ci$ agrees with experiment.
We have found that form $C1$ is incompatible with the phenomenology,
while both forms $C2$ and $C3$ are viable,
but only for the case of an inverted hierarchy.
Form $C2$ predicts $\cos{\delta} \gtrsim 0.67$
while form $C3$ predicts $\cos{\delta} \lesssim -0.67$;
both forms predict
$0.24 \le m_{\beta\beta} \left/ \sqrt{\Delta} \right. \le 0.34$;
furthermore,
these forms only work for $\sin^2{\theta_{12}} \gtrsim 0.325$,
\textit{cf.}~fig.~\ref{fig:C3}).

\subsection{Forms $B1$--$9$}

The mass matrix $M_B$ in the second eq.~(\ref{mk})
is of `Fritzsch type'~\cite{Fritzsch:1979zq}.
The exact diagonalization of a Fritzsch mass matrix
has been known for a long time~\cite{Li:1979zj}.
The use of Fritzsch-type mass matrices in the lepton sector
has been proposed before~\cite{xing}.

With $M_B$ the neutrino masses are not fixed.
In case n,
$m_1$ is the smallest neutrino mass,
$m_2 = \left( m_1^2 + \delta \right)^{1/2}$,
$m_3 = \left( m_1^2 + \Delta \right)^{1/2}$,
and
\bs
\ba
a &=& m_3 - m_2 + m_1,
\\
b &=& \sqrt{\frac{\left( m_3 - m_2 \right) \left( m_3 + m_1 \right)
\left( m_2 - m_1 \right)}{a}},
\\
d &=& \sqrt{\frac{m_3 m_2 m_1}{a}}.
\ea
\es
Then,
\ba
& & O_B =
\no & &
\left( \begin{array}{ccc}
- \sqrt{\frac{\left( m_3 + m_1 \right) \left( m_2 - m_1 \right) m_1}
{\left( m_3 - m_1 \right) a \left( m_2 + m_1 \right)}} &
\sqrt{\frac{\left( m_3 - m_2 \right) m_2 \left( m_2 - m_1 \right)}
{\left( m_3 + m_2 \right) a \left( m_2 + m_1 \right)}} &
\sqrt{\frac{m_3 \left( m_3 - m_2 \right) \left( m_3 + m_1 \right)}
{\left( m_3 + m_2 \right) \left( m_3 - m_1 \right) a}}
\\*[3mm]
\sqrt{\frac{\left( m_3 - m_2 \right) m_1}
{\left( m_3 - m_1 \right) \left( m_2 + m_1 \right)}} &
- \sqrt{\frac{\left( m_3 + m_1 \right) m_2}
{\left( m_3 + m_2 \right) \left( m_2 + m_1 \right)}} &
\sqrt{\frac{m_3 \left( m_2 - m_1 \right)}
{\left( m_3 + m_2 \right) \left( m_3 - m_1 \right)}}
\\*[3mm]
\sqrt{\frac{m_3 \left( m_3 - m_2 \right) m_2}
{\left( m_3 - m_1 \right) a \left( m_2 + m_1 \right)}} &
\sqrt{\frac{\left( m_3 + m_1 \right) m_3 m_1}
{\left( m_3 + m_2 \right) a \left( m_2 + m_1 \right)}} &
\sqrt{\frac{m_2 \left( m_2 - m_1 \right) m_1}
{\left( m_3 + m_2 \right) \left( m_3 - m_1 \right) a}}
\end{array} \right).
\no & &
\ea
If the matrix $\pmns$ has form $Bp$,
then one of its rows coincides,
in the moduli of its matrix elements,
with a row of $O_B$.
Considering the absolute values
of the matrix elements in the third column of $O_B$,
one finds that none of them can be equal to  either
$\sin{\theta_{23}} \cos{\theta_{13}}$
or $\cos{\theta_{23}} \cos{\theta_{13}}$---they are either too large
or too small for that.
On the other hand,
either $\left( O_B \right)_{23}$ or $\left( O_B \right)_{33}$
may coincide with $\sin{\theta_{13}}$.
However,
whenever this happens the other two matrix elements
in the corresponding row of $O_B$
are practically equal in absolute value,
which means that $\theta_{12}$ would be close to maximal,
contradicting phenomenology.
We thus conclude that the forms $B1$--$9$ for $\pmns$
are not viable in case n.

In case i,
$m_3$ is the smallest neutrino mass,
$m_1 = \left( m_3^2 + \Delta \right)^{1/2}$,
$m_2 = \left( m_3^2 + \Delta + \delta \right)^{1/2}$,
and
\bs
\ba
a &=& m_2 - m_1 + m_3,
\\
b &=& \sqrt{\frac{\left( m_2 - m_1 \right) \left( m_2 + m_3 \right)
\left( m_1 - m_3 \right)}{a}},
\\
d &=& \sqrt{\frac{m_2 m_1 m_3}{a}}.
\ea
\es
Moreover,
\ba
& & O_B =
\no & &
\left( \begin{array}{ccc}
\sqrt{\frac{\left( m_2 - m_1 \right) m_1 \left( m_1 - m_3 \right)}
{\left( m_2 + m_1 \right) a \left( m_1 + m_3 \right)}} &
\sqrt{\frac{m_2 \left( m_2 + m_3 \right) \left( m_2 - m_1 \right)}
{\left( m_2 + m_1 \right) \left( m_2 - m_3 \right) a}} &
- \sqrt{\frac{\left( m_2 + m_3 \right) \left( m_1 - m_3 \right) m_3}
{\left( m_2 - m_3 \right) a \left( m_1 + m_3 \right)}}
\\*[3mm]
- \sqrt{\frac{\left( m_2 + m_3 \right) m_1}
{\left( m_2 + m_1 \right) \left( m_1 + m_3 \right)}} &
\sqrt{\frac{m_2 \left( m_1 - m_3 \right)}
{\left( m_2 + m_1 \right) \left( m_2 - m_3 \right)}} &
\sqrt{\frac{\left( m_2 - m_1 \right) m_3}
{\left( m_2 - m_3 \right) \left( m_1 + m_3 \right)}}
\\*[3mm]
\sqrt{\frac{\left( m_2 + m_3 \right) m_2 m_3}
{\left( m_2 + m_1 \right) a \left( m_1 + m_3 \right)}} &
\sqrt{\frac{m_1 \left( m_1 - m_3 \right) m_3}
{\left( m_2 + m_1 \right) \left( m_2 - m_3 \right) a}} &
\sqrt{\frac{m_2 \left( m_2 - m_1 \right) m_1}
{\left( m_2 - m_3 \right) a \left( m_1 + m_3 \right)}}
\end{array} \right).
\no & &
\ea
In this case one finds that either the first or the third row of $O_B$
are suitable to fit either the second or the third row of $\pmns$;
this means that the forms $B2$,
$B3$, $B8$, and $B9$ of $\pmns$ are viable.

\begin{figure}[hbt]
\centering
\includegraphics[width=10cm,angle=0]{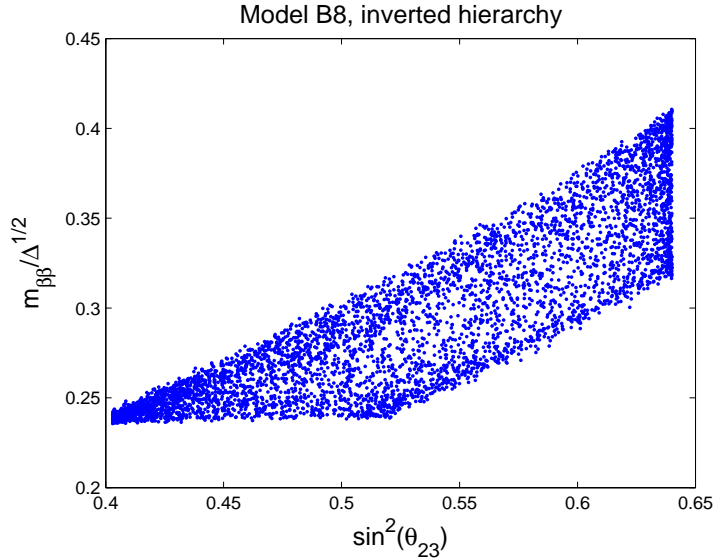}
\caption{$m_{\beta \beta} \left/ \sqrt{\Delta}\right.$
\textit{vs} $\sin^2\theta_{23}$ in form $B8$.
The numeric scan was made by using the $3\, \sigma$ data
of ref.~\cite{Forero:2014bxa}.}
\label{fig:B8mbb}
\end{figure}

\begin{figure}[hbt]
\centering
\includegraphics[width=10cm,angle=0]{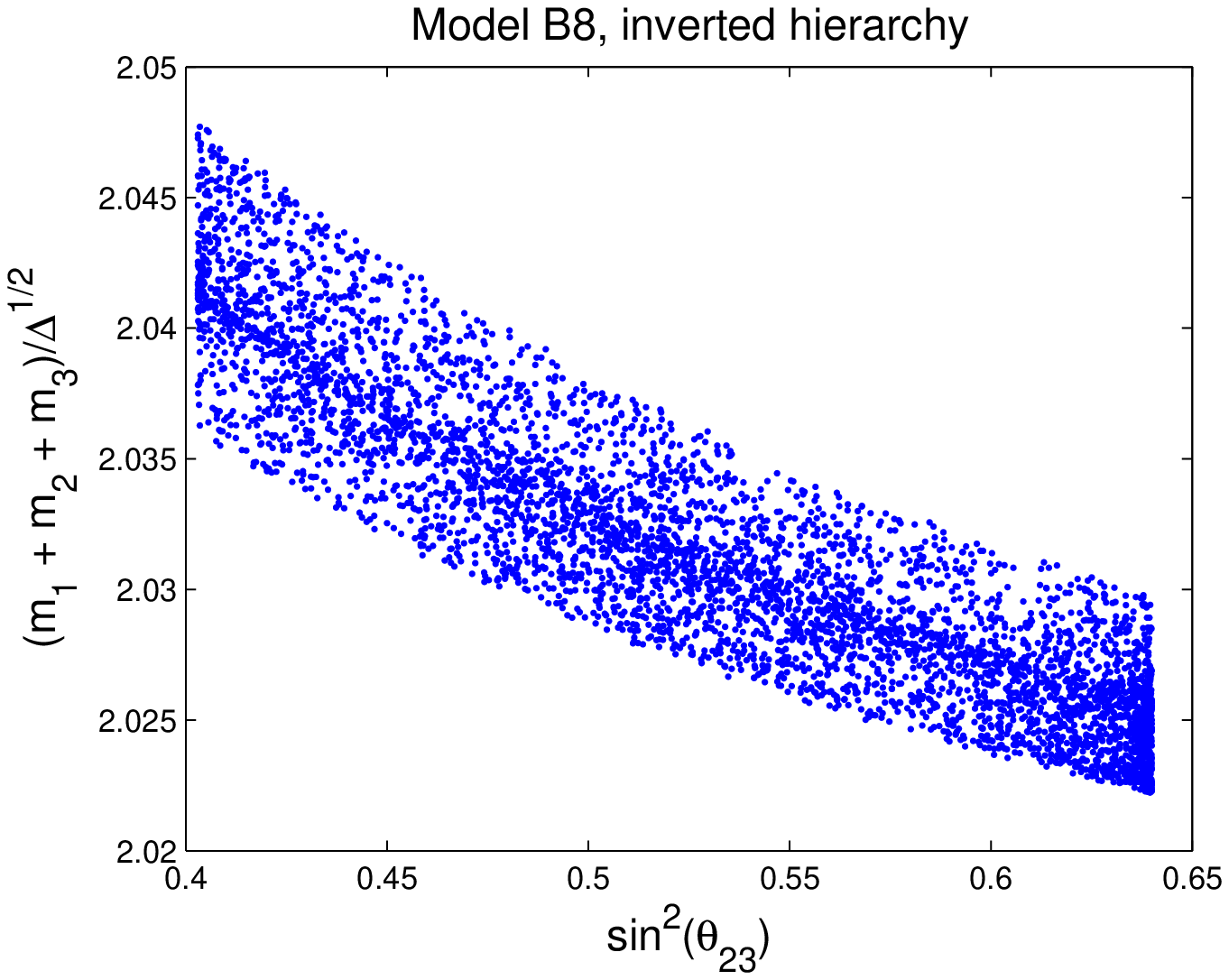}
\caption{$\left( m_1 + m_2 + m_3 \right) \left/ \sqrt{\Delta}\right.$
\textit{vs} $\sin^2\theta_{23}$ for form $B8$.
The numeric scan was made by using the $3\, \sigma$ data
of ref.~\cite{Forero:2014bxa}.}
\label{fig:B8msum}
\end{figure}

Forms $B2$ and $B8$ predict a positive $\cos{\delta}$:
$\cos{\delta} > 0.37$ for form $B2$
and $\cos{\delta} > 0.58$ for form $B8$.
They both privilege higher-than-average $\theta_{12}$ and $\theta_{23}$---both
angles are not allowed to be simultaneously below their best-fit values.
The overall scale of the neutrino masses is given by
$2.023 \le \left( m_1 + m_2 + m_3 \right) \left/ \sqrt{\Delta} \right.
\le 2.050\ (2.047)$ for form $B2$ ($B8$);
neutrinoless $\beta \beta$ decay is governed by
$0.24 \le m_{\beta \beta} \left/ \sqrt{\Delta} \right. \le 0.47, 0.41$
for forms $B2$ and $B8$,
respectively.
We also find some broad correlation between these mass ratios
and $\sin^2{\theta_{23}}$,
as may be seen in figs.~\ref{fig:B8mbb} and~\ref{fig:B8msum}
for the form $B8$.
Similar correlations occur for form $B2$,
but there the variation is opposite:
$\left( m_1 + m_2 + m_3 \right) \left/ \sqrt{\Delta}\right.$
increases with $\sin^2{\theta_{23}}$
and $m_{\beta \beta} \left/ \sqrt{\Delta}\right.$ decreases.

Forms $B3$ and $B9$ of $\pmns$ are similar to forms $B2$ and $B8$,
respectively,
with the interchange $\nu_\mu \leftrightarrow \nu_\tau$.
Therefore the predictions are broadly similar,
only $\cos{\delta}$ is predicted to be negative instead of positive
and $\theta_{23}$ is expected to be small,
\textit{viz}.\ in the first octant,
rather than large.
Again,
there are hints of correlations of mass parameters with
$\sin^2{\theta_{23}}$,
similar to those found for models $B2$ and $B8$,
but they now appear reversed:
for model $B3$,
$\left( m_1 + m_2 + m_3 \right) \left/ \sqrt{\Delta}\right.$ decreases
with $\sin^2{\theta_{23}}$ and $m_{\beta \beta} \left/ \sqrt{\Delta}\right.$
increases
(the exact opposite of what occurred for model $B2$).
Likewise,
the behaviour of these correlations for model $B9$ is the
opposite of what occurred for model $B8$.

\section{Synopsis}

In this paper we have found that there are six (5, 2) textures
that are still viable:
they are listed in eqs.~(\ref{mxopd}),
wherein $H_{12} = 0$ is the consequence of one of the $M_\ell$ textures
in eqs.~(\ref{viogp}).
As for (4, 3) textures,
there are eight of them which agree with present-day phenomenology;
the corresponding forms the lepton mixing matrix are given
in eqs.~(\ref{b2}),
(\ref{b3}),
(\ref{b8}),
(\ref{b9}),
(\ref{155}),
(\ref{16}),
(\ref{80}),
and (\ref{90});
the corresponding textures of $M_\nu$ are those
in eqs.~(\ref{d4})--(\ref{d6}).\footnote{Some of our textures
had already been presented,
although from a different vantage point,
in ref.~\cite{Barbieri:2003qd}.
\emph{All}\/ of our textures have been independently derived
in a recent paper~\cite{Ludl:2014axa}.}

Even though there such a large variety of viable textures,
the same cannot be said about the ensuing predictions,
which are broadly similar for all of them:
all the viable textures only tolerate
\begin{itemize}
\item an inverted neutrino mass spectrum,
\item an overall scale of the neutrino masses given by
$\left( m_1 + m_2 + m_3 \right) \left/ \sqrt{\Delta} \right.$
in the range $\left[ 2.0,\ 2.1 \right]$,
\item $\cos{\delta}$ far away from 0,
\textit{i.e.}\ close to either $+1$ or $-1$,
and
\item neutrinoless double-beta decay governed
by $m_{\beta \beta} \left/ \sqrt{\Delta} \right. \in \left[ 0.24,\  0.48 \right]$.
\end{itemize}
%
We thus conclude that texture-zero models
of the (5, 2) and (4, 3) varieties
are quite monotonous in their predictive power.

\vspace*{5mm}

\paragraph{Acknowledgements:}
The work of PMF is supported by
the Portuguese Foundation for Science and Technology (FCT)
through the projects PEst-OE/FIS/UI0618/2011 and PTDC/FIS/117951/2010,
and through the FP7 Reintegration Grant PERG08-GA-2010-277025.
The work of LL is supported through
the Marie Curie Initial Training Network ``UNILHC'' PITN-GA-2009-237920
and also through the projects PEst-OE-FIS-UI0777-2013,
PTDC/FIS-NUC/0548-2012,
and CERN-FP-123580-2011 of FCT.

\end{document}